\documentclass[12pt]{article}
\usepackage{amsfonts}
\usepackage{bbding}
\usepackage{mathrsfs}
\usepackage{amssymb}
\usepackage{amsmath}
\usepackage{graphicx}
\usepackage{color}
\usepackage{epsfig}

\def\p{\partial}

\def\a{\alpha}
\def\b{\beta}

\def\D{\Delta}

\def\g{\gamma}

\def\l{\lambda}

\def\O{\Omega}

\def\s{\sigma}

\def\td{\tilde}

\def\defn{\stackrel{\triangle}{=}}

\def\td #1{{\tilde #1}}

\newtheorem{prop}{Proposition}[section]

\begin{document}
\title{\bf CVA and FVA to Derivatives Trades Collateralized by Cash}
\author{
Lixin Wu \thanks{Part of the results of this paper have been presented in the World Congress of Bachelier Finance Society, June,
2012, Sydney; The First Hong Kong-Shanghai Workshop on Quantitative Finance and Risk Management, September, 
2012, Shanghai; Mathematical Finance Seminar in Imperial College, October, 
2012, London; the Workshop Stochastic Control and Financial Applications by AMSS-PolyU Joint Research Institute, December, 
2012, Hong Kong; and the First Asian Quantitative Finance Conference, January 2013, Singapore. I would like to thank Chunhong Li for computational assistance, and Damiano Brigo, Fabio Mecurio and Eckhard Platen for their helpful comments. All remaining errors are mine.}  \\
Department of Mathematics\\
University of Science and Technology \\
Clear Water Bay, Kowloon \\
Hong Kong
 \\
 \\
\date{January 18, 2013}
}
\maketitle

\newpage
\begin{abstract}
In this article, we combine replication pricing with expectation pricing for derivative trades that are partially collateralized
by cash. The derivatives are replicated by underlying assets and cash, using repurchasing agreement (repo) and margining,
which incur funding costs.
We derive a partial
differential equation (PDE) for the derivatives price, obtain and decompose its solution into the risk-free
value of the derivative plus credit valuation adjustment (CVA) and funding valuation adjustment (FVA).
For most derivatives, as
we shall show, CVAs can be evaluated analytically or semi-analytically, while FVAs, as well as the derivatives values, will have to be solved recursively through numerical procedures due to their interdependence.
In numerical demonstrations, continuous and discrete margin revisions are considered, respectively, for an equity call option and a  vanilla interest-rate swaps.

\end{abstract}
%



\newpage

\section{Introduction}
One of the major consequences of the 2007-08 financial tsunami is the disappearance of the boundary between market risk and credit risk. After the tsunami, even popular vanilla trades on market risks are not immune from counterparty default risks, and accompanied
with which also come the funding risks, meaning the higher and uncertain funding costs.
To cope with the counterparty default risk, financial institutions have been increasingly adopting the practice of collateralization, creating the differentiation between the so-called CSA and non-CSA trades\footnote{CSA is for Credit Support Annex, a legal document of International Swap and Derivatives Association (ISDA) that regulates collateral posting. Non-CSA trades are also regulated by the ISDA master agreement.}.
Since the financial tsunami, pricing and managing both CSA and non-CSA derivative trades has been a major focus in the industry.
Today, the use of credit valuation adjustment (CVA) has become the market standard to account for the counterparty default risks, yet
discussions and research are still ongoing regarding the so-called funding valuation adjustment (FVA) for funding risks.
Since credit and funding risks are intimately related, a recent trend is to evaluate these two kinds of risks consistently under a unified framework. In this article we present such a framework based on arbitrage pricing theory for the consistent evaluation of bilateral CVA and FVA for trades collateralized through, in particular, margining.


Early literatures on counterparty risks may be traced back to Sorensen and
Bollier (1994), where the pricing of interest-rate swaps subject to counterparty default risks is considered. The same problem is also studied in Duffie and Huang (1996), where any payout is discounted using the discount rate of the paying party. Their model was later extended by Huge and Lando (1999) to explicitly account for the credit ratings of the counterparties. Over the years there has been more research, including Bielecki and Rutkowski (2001), Canabarro and Duffie (2004), and Picoult (2005). Gradually during the same period of time, the notion of credit valuation adjustment was established in industry. The effects of various new market features on CVA have also been studied, including ``netting agreements", which allow multiple obligations to be consolidated into a single one upon a default, by Brigo and Masetti (2005) and also by Pyktin and Zhu (2007), and wrong-way risk by Brigo and Pallavicini (2007).
As an important feature, collaterals was considered, among others, by Cherubini (2005) in the evaluation of CVA for basic
products, and then in Li and Tang (2007) for general derivatives. Alavian {\it et al.} (2008)
discuss minimum transfer amounts and collateral thresholds,
and provide model independent formulas for the counterparty exposure.
Assefa {\it et al.} (2009) introduce
a model for the collateral process without accounting for minimum
transfer amounts and collateral thresholds. Other than to the usual equity and interest-rate derivatives, the notion of CVA has
also been extended to other asset classes, including commodity derivatives by Brigo and Bakkar (2009) and credit derivatives by Brigo and Chourdakis (2009).

Studies of funding valuation adjustment has become a focus only in recent years. In a classical Black-Scholes framework, Piterbarg (2010) derives replication pricing of derivatives under collateralization but without default risk. Funded replication pricing is also considered by Fries (2010) using the technique for pricing quantos, for which cash flows are indexed in one currency but paid in another. Fujii {\it et al.} (2010) analyze implications of currency risk for collateral modeling.
For simple products like zero-coupon bonds or loans, Morini and Prampolini (2011) and Castagna (2011) investigate essential features of funding costs in the presence of default risk.

It is not hard to see that CVA and FVA are interrelated. While collaterals
mitigate potential losses upon the counterparty defaults and thus reduce the value of CVA, they may incur additional funding cost to the posting party, which should be properly accounted for in the valuation process of derivatives.
A current trend, seen for example in Burgard and Kjaer (2011), Pallavicini {\it et al.} (2011), Lu and Juan (2011) and Castagna (2011), is to study CVA and FVA in a consistent framework.
Pallavicini {\it et al.} (2011) derive a nonlinear and recursive equation for the derivative
value subject to counterparty and funding risks, with or without collaterals, and resort to an iterative method for its solution. A similar equation is also obtained by Cr\'epey (2011). For un-collateralized trades with
default payments governed by the 2002 ISDA master agreement, Burgard and Kjaer (2011) present a framework of replication pricing using underlying share and zero-coupon bonds. Wu (2012) replaces the zero-coupon bonds by credit default swaps in the replication arguments and derives explicit formulae for CVA and FVA under stochastic interest rates.

In this article, we will price derivatives trades that are partially collateralized by cash through unsecured borrowing, so-called
margining, with collateral posting subject to a threshold value and a minimum transfer amount. We start with the issuer's hedging portfolio,
which consists of the underlying asset and cash accounts, and define the fair value of a derivative to be the premium payment that
makes the expected present values of the cash flows from the two parties equal. A direct application of the Ito's lemma to the price
dynamics of the derivatives subject
to counterparty default and funding risks enables us to formulate CVA and FVA under a general context, including general asset price dynamics, stochastic interest rates, stochastic hazard rates and stochastic recovery rates. In a more transparent way, our results demonstrate the interdependence
between the derivatives value and its FVA.
For most derivatives, as
we shall show, CVA can be evaluated analytically or semi-analytically, while FVA, as well as the derivatives value, will have to be solved recursively through numerical procedures due to their interdependence.


The combined approach of pricing has several distinct advantages. First, by including the cost of margining to the buyer's portfolio, we obtain a unique price for the derivative,
regardless which party, the seller or the buyer, hedges the market risks.
Second, for the first time, we are able to identify and
evaluate all adjustment terms in generality. Third, the formulae for CVA and FVA offer clear insights to the active management of counterparty risks and funding risks, allowing them to be marked to market and actively hedged. Finally, the combined approach also
has the potential to price other features like
re-hypothecation, when collaterals are lent out for funding, and wrong-way risk, when exposure to a counterparty is adversely correlated with the credit quality of that counterparty. These features are however left for future research.

This article is organized as follows. In section 2 we start with replication pricing of an equity derivative and then switch to the martingale approach for pricing. In section 3 we study the actual evaluation of the adjustment terms, delivering closed-form formulae for CVA while describing numerical methods for FVA. A numerical example with continuous margin revision is presented. In section 4 we consider the pricing of interest-rate swaps under collateral posting through discrete margin revision, and again deliver closed-form solution to CVA. Finally, we conclude in section 5. Some technical details are placed in the appendix.

\section{Arbitrage Pricing}
\subsection{Default payments}
Consider the pricing of a derivative trade, with or without collateral, between two defaultable parties, $B$ a bank (or seller) and $C$ a counterparty (or buyer). Prior to the default of either party, the derivative generates the cash flows according to the contract. In case of a default, the default payment can be described in two ways, depending on whether or not there is a collateral.
Let $V(t)$ (or $V_t$) denote the pre-default value of the derivative to party $C$,
which is left continuous with right limit, so-called a c$\grave{a}$gl$\grave{a}$d function,
$M(t)$ be the MTM value of the derivative, both are seen from the eyes of the buyer. When there is no
collateral, the defaulted party pays the surviving party, according to the
ISDA master agreement,
the recovered value of its positive mark-to-market (MTM) value obtained through a dealer poll mechanism:
\begin{equation}\label{e2010}
\begin{array}{ll}
V(\tau_B+)=R_BM^+(\tau_B)+M^-(\tau_B), &\mbox{if $B$ defaults,} \\
V(\tau_C+)=M^+(\tau_C)+R_CM^-(\tau_C), &\mbox{if $C$ defaults,}
\end{array}
\end{equation}
where $R_B$ and $R_C$ are recovery rates, with $0\leq R_B,R_C\leq 1$. The polling mechanism usually produces an MTM value close to the risk-free value of the derivative. The other possibility for $M(\tau)$ is the pre-default value.

When there is a collateral, the
surviving party will keep the collateral, or part of the collateral if it is worth more than the MTM value of the derivative. Let $c(t)$ or $c_t$ be the value of the collateral, then
the payment upon a default or the post-default value of the derivative can be expressed as
\begin{equation}\label{e2011}
\begin{array}{ll}
V(\tau_B+)=\min\{c^+(\tau_B),M^+(\tau_B)\}+M^-(\tau_B), &\mbox{if $B$ defaults,} \\
V(\tau_C+)=\max\{c^-(\tau_C),M^-(\tau_C)\}+M^+(\tau_C), &\mbox{if $C$ defaults,}
\end{array}
\end{equation}
where $\tau_B$ and $\tau_C$ are the default time of the two parties,
$c^+(\tau),c^-(\tau)$ and $M^+(\tau), M^-(\tau)$ are defined according to $f^+(\tau)=\max\{f(\tau),0\}$ and $f^-(\tau)=\min\{f(\tau),0\}$, respectively.
The collateral can be either cash or other securities.
When $c(t)>0$, the collateral is posted by party $B$ and placed under the custody of party $C$. When $c(t)<0$, it is the opposite. Normally, there are $c(t)\geq 0$ for
$M(t)>0$, and $c(t)\leq 0$ for $M(t)<0$.
When the collateral is in the form of cash and is withdrawn from a margin account, we call it margining.


According to ISDA's master agreement, the value of cash collateral can be expressed as
\begin{equation}\label{e2012}
\begin{array}{ll}
c^+(t)&=\left(M(t)-H+X\right)1_{\{M(t)\geq H\}}, \\
c^-(t)&=\left(M(t)+H-X\right)1_{\{M(t)\leq -H\}},
\end{array}
\end{equation}
where $H$ is the threshold for collateral posting and $X$ is the minimum transfer amount, both can be time dependent, with $H\geq X\geq 0$.

Note that (\ref{e2010}) can be deduced from (\ref{e2011}), for
\begin{equation}\label{e2013}
\begin{array}{ll}
c^+(t)&=R_BM^+(t), \\
c^-(t)&=R_CM^-(t).
\end{array}
\end{equation}
Therefore, we will use (\ref{e2011}) as the general formulation of default payments for both un-collateralized
and collateralized trades.

\subsection{Replication and expectation pricing}
Derivatives pricing in incomplete markets can be done as follows. First, we hedge the risks, as much as we can, using liquid instruments. Then, we assign risk premium to residual risks, if any, exposed by the hedged portfolio. These two steps will give rise to a governing equation
for the value function. Funding costs arise in hedging as well as margining. The key to formulate the funding costs is to characterize the evolution of the margin accounts.

Without loss of generality, we consider the pricing of an equity derivative, which is allowed to switch between asset ($V_t>0$) and liability ($V_t<0$) over its life. Suppose at time $t=0$ the bank ``sells" a derivative to counterparty $C$ at price $V_0$.
To hedge off risks, at any time $t\geq 0$ the issuer $B$ forms a
portfolio using the underlying share and cash:
\[
\Pi_B(t)=\a_SS_t+[(\b_B(t)-c_t)-\a_SS_t+c_t],\]
where
$S_t$ is the value of the underlying share, $\a_S$, yet to be chosen, is the number of units of the underlying security for hedging, and $\b_B(t)$ is the balance of the margin account before the actions of hedging and collateral posting.
The three terms within the square brackets represent three different cash accounts.
The first term is the balance of the margin account, which earns risk-free rate or costs risk-free rate plus a spread, respectively;
the second term is the capital generated by ``repoing in" or ``repoing out" the underlying for hedging purpose,
and this piece of capital earns or costs the repo rate (corresponding to the underlying share), which normally has a spread over the risk-free rate;
and the third term is the value of the
collateral held by the counterparty (when $c_t>0$) or posted by the counterparty (when $c_t<0$), which however
earns or costs only the risk-free rate. Initially, the value of the portfolio is
\[
\Pi_B(0)=V_0^+\geq 0,\]
and it evolves according to
\begin{equation}
\begin{split}\label{e2090}
d\Pi_B(t)=r_t(\b_B(t)&-c_t)dt+x_B[\b_B(t)-c_t]^-dt \\
&+\a_SdS_t-(r_t+\l_S)\a_SS_tdt+r_tc_tdt,
\end{split}
\end{equation}
where $r_t$ is the risk-free spot rate, $\l_S$ is the repo spread corresponding to the underlying share, and $x_B(t)\geq 0$ is the total funding spread to $B$ for unsecured borrowing.

The buyer, meanwhile, also has a portfolio, which consists of the derivative and cash for funding purpose:
\[
\Pi_C(t)=V_t+\b_C(t).\]
When $V_t<0$, the derivative becomes a liability to the buyer, who then is liable to post cash collateral of value $-c_t\geq 0$ to $B$,
resulting in the following aggregated balance of his cash accounts:
\[\b_C(t)=[\b_C(t)+c_t]- c_t,\]
where the first term is balance in the buyer's margin account, and the second term is the value of the collateral posted to $B$. The aggregated return of $C$'s cash accounts is described by
\begin{equation}
\begin{split}\label{e20921}
d\b_C(t)=r_t(\b_C(t)+c_t)dt+x_C[\b_C(t)+c_t]^-dt-r_tc_tdt,
\end{split}
\end{equation}
subject to initial value
\[
\b_C(0)=-V_0^-\geq 0,\]
where in (\ref{e20921}) $x_C(t)$ is the total funding spread to $C$ for unsecured borrowing.

We model the uncertain market by the probability space $(\O,{\cal G},{\cal G}_t,{\Bbb P})$, where ${\Bbb P}$ is the physical measure, ${\cal G}_t$ represents all market information up to time $t$. We can write ${\cal G}_t={\cal F}_t\vee {\cal H}_t$, where ${\cal F}_t$ is the usual filtration which contains all market information except defaults while ${\cal H}_t=\s(\{\tau_B\leq u\}\vee\{\tau_C\leq u\}:u\leq t)$ carries
only the information of default.
To avoid unnecessary complication, we assume no correlation between the derivatives value and the credit worthiness of the two counterparties.
For notational simplicity we assume throughout this article that there is no
default by the current moment $t=0$, i.e., $\tau>0$.

For efficiency of presentation, we will work directly under the risk-neutral measure ${\Bbb Q}\sim {\Bbb P}$, which is the martingale measure corresponding to the numeraire of money market account,
\begin{equation}
\begin{split}\label{e2060}
B_t=e^{\int^t_0 r_udu},
\end{split}
\end{equation}
and can be constructed by usual arguments for change of measure\footnote{Note that ${\Bbb Q}$ can be uniquely determined if, in addition to the underlying share, default risks and recovery risks can also be traded in the markets, through, for example, credit default swaps and recovery swaps.}.
The risk-neutral price dynamics of the underlying asset is
\begin{equation*}
\begin{split}
dS_t&=S_t\left[(r_t-q_t)dt+\s_S(t) dW_t\right], \\
\end{split}
\end{equation*}
where
$W_t$ is a one-dimensional Brownian motion under ${\Bbb Q}$,
$\s_S$ is the percentage volatility of $S_t$, and
$q_t$ is the dividend yield of the share. 

To avoid potential complications brought by the possibly stochastic interest rate, we will work with asset prices discounted by money market account:
\[
\hat A_t={A_t\over B_t},\]
where $A_t$ represents the cum-dividend price of any tradeable assets.
It can be easily verified that discount price of the cum-dividend share,
\[
\hat S_t={S_te^{\int^t_0q_sds}\over B_t},\]
has the following risk-neutral dynamics:
\begin{equation*}
d\hat S_t=\hat S_t\s_S(t) dW_t.
\end{equation*}
The discount value of the seller's hedging portfolio then
evolves according to
\begin{equation}
\begin{split}\label{e2065}
d\hat \Pi_B(t)=\a_S\hat S_t\s_S dW_t-\a_S\l_S\hat S_tdt+x_B[\hat \b_B(t)-\hat c_t]^-dt.
\end{split}
\end{equation}
On the other hand, the discount value of the buyer's portfolio is
\[
\hat \Pi_C(t)=\hat V_t+\hat \b_C(t).\]
According to the Ito's lemma,
\begin{equation*}
\begin{split}
d\hat V_t=&\left(\p_t\hat V_t+{1\over 2}\hat S^2_t\s_S^2\p^2_S\hat V_t\right)dt+\p_S\hat V_t\hat S_t\s_S dW_t+\D\hat V_BdJ_B+\Delta\hat V_CdJ_C,
\end{split}
\end{equation*}
where $J_B$ and $J_C$ are two independent Poisson processes that jump from 0 to 1 with risk-neutral intensities $\l_B$ and $\l_C$,
$\D\hat V_B$ and $\D\hat V_C$ stand for the jump sizes upon the first default of $B$ and $C$, while the aggregated discount value of the buyer's cash accounts grows according to
\[
d\hat \b_C(t)=x_C[\hat \b_C(t)+\hat c_t]^-dt.\]
It then follows that
\begin{equation}
\begin{split}\label{e2075}
d\hat\Pi_C(t)=&\left(\p_t\hat V_t+{1\over 2}\hat S^2_t\s_S^2\p^2_S\hat V_t\right)dt+\p_S\hat V_t\hat S_t\s_S dW_t+\D\hat V_BdJ_B+\D\hat V_CdJ_C \\
&+x_C[\hat \b_C(t)+\hat c_t]^-dt.
\end{split}
\end{equation}
Subtracting (\ref{e2065}) from (\ref{e2075}) we obtain the difference of returns between the seller's and buyer's portfolios:
\begin{equation}
\begin{split}\label{e2105}
d\hat\Pi_C(t)-d\hat \Pi_B(t)=&\left(\p_t\hat V_t+{1\over 2}\hat S^2_t\s_S^2\p^2_S\hat V_t\right)dt+\D\hat V_BdJ_B+\D\hat V_CdJ_C \\
&+(\p_{\hat S}\hat V_t-\a_S)\hat S_t\s_S dW_t \\
+\l_S\a_Sdt&-x_B[\hat \b_B(t)-\hat c_t]^-dt+x_C[\hat \b_C(t)+\hat c_t]^-dt.
\end{split}
\end{equation}
To hedge against the diffusion risks, $B$ should take
\begin{equation*}
\begin{split}
\a_S=\p_{\hat S} \hat V_t.
\end{split}
\end{equation*}
Suppose the residual jump risks can be hedged by using credit default swaps (CDS), then, because upon
entry a CDS has no value, we assign no risk premium to the jump risks and thus can set the ${\Bbb Q}$ expectation of (\ref{e2105}) to zero. After rearranging, we obtain the governing PDE for the value of the derivative:
\begin{equation}
\begin{split}\label{e2110}
\p_t\hat V_t+{1\over 2}\s_S^2\hat S^2\p^2_{\hat S}\hat V_t =&-\l_BE_t[\D\hat V_B]-\l_CE_t[\D\hat V_C]\\
&-\l_S\hat S\p_{\hat S}\hat V_t+x_B[\hat \b_B(t)-\hat c_t]^--x_C[\hat \b_C(t)+\hat c_t]^-,
\end{split}
\end{equation}
where $\l_B$ and $\l_C$ are the respective risk-neutral intensities of default for $B$ and $C$, and the conditional expectation is defined by
\[
E_t[X]:=E^Q[X|{\cal F}_t\vee \{\tau>t\}].\]
Equation (\ref{e2110}) is subject to (\ref{e2011}), the payment upon default, or to the usual terminal condition of derivatives payoff at maturity. 
Note that (\ref{e2110}) is valid for both deterministic and stochastic spot interest rates, and it will become a lot more complex if instead spot prices are used under stochastic interest rates. 

Using the Feynman-Kac formula, we obtain the following expression of the solution to (\ref{e2110}):
\begin{equation}
\begin{split}\label{e2120}
\hat V_0=E^{Q}_0[\hat V_{T\wedge\tau}]
    &+E^{Q}_0\left[\int^{T\wedge\tau}_0\left(\l_BE_u[\D\hat V_B]+\l_CE_u[\D\hat V_C]\right.\right.  \\
    &\left.\left. +\l_S\hat S_u\p_{\hat S}\hat V_u-x_B[\hat \b_B(u)-\hat c_u]^-+x_C[\hat \b_C(u)+\hat c_u]^-\right)du\right],
\end{split}
\end{equation}
where $\tau=\tau_B\wedge\tau_C$.
For actual valuation, (\ref{e2120}) is not too useful due to the presence of the path-dependent integral with stochastic upper limit. Additional efforts are required to solve for $\hat V_0$.

We now derive an alternative Feynman-Kac formula for the process of the value function.
Combining the diffusion and the first jump of the value function, we have
\begin{equation*}
    \begin{split}
        [1_{\tau>T}&\hat V_T+1_{\tau\leq T}\hat V(\tau+)]-\hat V_0 \\
        =&\int^{T\wedge\tau}_0\left[\left(\p_u\hat V_u+{1\over 2}\hat S_u^2\s_S^2\p^2_{\hat S}\hat V_u\right)du+\hat S_u\p_{\hat S}\hat V_u\s_S d{W}_u+\D\hat V_udJ_u\right].
     \end{split}
\end{equation*}
Conditional on ${\cal F}_0\vee\{\tau>0\}$, we take expectation on both sides of the above equation, then, by making use of the equation (\ref{e2110}), we obtain
\begin{equation*}
    \begin{split}
        &E^{Q}_0[1_{\tau>T}\hat V_T]+E^{Q}_0[1_{\tau\leq T}\hat V(\tau+)]-\hat V_0 \\
        =&E^{Q}_0\left[\int^{T\wedge\tau}_0\left(-\l_S\hat S_u\p_{\hat S}\hat V_u-x_B[\hat \b_B(u)-\hat c_u]^-+x_C[\hat \b_C(u)+\hat c_u]^-\right)du\right].
     \end{split}
\end{equation*}
Rearranging, we end up with
\begin{prop} 
The solution to (\ref{e2110}) is
\begin{equation}
\begin{split}\label{e2150}
\hat V_0=&E^{Q}_0[1_{\{\tau>T\}}\hat V_T]+E^{Q}_0[1_{\{\tau\leq T\}}\hat V(\tau+)] \\
&+E^{Q}_0\left[\int^{T\wedge\tau}_0\left(\l_S\hat S_u\p_{\hat S}\hat V_u-x_B[\hat \b_B(u)-\hat c_u]^-+x_C[\hat \b_C(u)+\hat c_u]^-\right)du\right]\quad\Box
\end{split}
\end{equation}
\end{prop}

Next, we will build the connection between the fair value of the derivative and its risk-free counterpart,
$$\hat V_e(0)=\hat V_e(\hat S_0,0)\defn E^{Q}_0[\hat V_T].$$
Obviously there is
\begin{equation}
\begin{split}\label{e2160}
E^{Q}_0[1_{\{\tau>T\}}\hat V_T]
    &=E^{Q}_0[\hat V_T]-E^{Q}_0[1_{\{\tau\leq T\}}\hat V_T].
\end{split}
\end{equation}
By the {\it tower law\/}, we have
\begin{equation}\label{e2170}
    \begin{split}
        E^{Q}_0[1_{\{\tau\leq T\}}\hat V_T]
        =& E^{Q}_0[1_{\{\tau\leq T\}}E^{Q}[\hat V_T|{\cal G}_\tau]] \\
        =& E^{Q}_0[1_{\{\tau\leq T\}}E^{Q}[\hat V_T|{\cal F}_\tau]] \\
        =& E^{Q}_0[1_{\{\tau\leq T\}}\hat V_e(\tau)].
     \end{split}
\end{equation}
Substituting (\ref{e2160}) and (\ref{e2170}) back to (\ref{e2150}), we obtain
\begin{equation*}
\begin{split}
\hat V_0&=E^{Q}_0[\hat V_T]+E^{Q}_0[1_{\{\tau\leq T\}}(\hat V(\tau+)-\hat V_e(\tau))] \\
&+E^{Q}_0\left[\int^{T\wedge\tau}_0\left(\l_S\hat S_u\p_S\hat V_u-x_B[\hat \b_B(u)-\hat c_u]^-+x_C[\hat \b_C(u)+\hat c_u]^-\right)du\right].
\end{split}
\end{equation*}
By distinguishing between $\tau=\tau_B$ and $\tau=\tau_C$ and noticing $\hat V_0=V_0$ and $\hat V_e(0)=V_e(0)$, we arrive at the key result of this article.
\begin{prop}\label{p22}
The value of a derivative under counterparty default and funding risks is given by
\begin{equation}
\begin{split}\label{e2190}
V_0&=V_e(0)+E^{Q}_0[1_{\{\tau=\tau_B\leq T\}}(\hat V(\tau_B+)-\hat V_e(\tau_B))] \\
&\quad\quad\quad+E^{Q}_0[1_{\{\tau=\tau_C\leq T\}}(\hat V(\tau_C+)-\hat V_e(\tau_C))] \\
+E^{Q}_0&\left[\int^{T\wedge\tau}_0\left(\l_S\hat S_u\D_u-x_B[\hat \b_B(u)-\hat c_u]^-+x_C[\hat \b_C(u)+\hat c_u]^-\right)du\right],
\end{split}
\end{equation}
where $\D_u=\p_{\hat S}\hat V_u$, and $\hat \b_B(t)$ and $\hat \b_C(t)$ satisfy the following evolution equations
\begin{equation}
\begin{split}\label{e2197}
d\hat \b_B(t)&=\a_S\hat S_t\s_S dW_t-\a_S\l_S\hat S_tdt+x_B[\hat \b_B(t)-\hat c_t]^-dt,\quad \b_B(0)=V^+_0, \\
d\hat \b_C(t)&=x_C[\hat \b_C(t)+\hat c_t]^-dt, \quad \b_C(0)=-V^-_0.
\end{split}
\end{equation}
The last three terms of (\ref{e2190}) represent respectively the CVA for the default risk of $B$ and $C$ and FVA to the derivatives value.\footnote{The combined CVA for the default risks of $B$ and $C$ is also called bilateral CVA. To either party, the CVA for a party's own default risk is also called debit valuation adjustment (DVA).}$\quad\Box$
\end{prop}

It can be seen that funding cost terms due to margining depend on $V_0$, the initial premium of the derivative, so (\ref{e2190}) only implicitly defines $V_0$, which in general will have to be
solved numerically with some methods of recursion.
While the formulation of CVAs in (\ref{e2190}) is similar to some existing results in the literature, e.g. Gregory (2009) and Brigo and Capponi (2010), the formulation of FVA is more specific and transparent than existing results in, e.g. Pallavicini {\it et al.} (2011) and Cr\'epey (2011).

We have a few more remarks. First, Proposition 2 remains valid for stochastic default intensities and funding spreads.
Second, formula (\ref{e2190}) can be generalized to pricing a derivative on multiple underlying assets, simply by replacing the funding cost for a single repo by that for a portfolio of repos on different underlying assets. Finally, the proposition can be naturally generalized to pricing a portfolio of derivatives with close-out netting agreements between a single counterparty (see e.g. Pykin and Zhu (2007)), or a derivative with multiple cash flows, including, for example,
caps/floors, interest-rate swaps, credit default swaps and etc.

We may decompose a firm's funding spread for unsecured borrowing into two components: firm-specific funding spreads and market-wide funding spread, the latter may arise during a credit crunch. Arguably, the firm-specific funding spread is approximately the credit default swap rate on the firm. Thus we may write
\[x_B=\l_BE_t[L_B]+\l_M,\quad\mbox{and}\quad x_C=\l_CE_t[L_C]+\l_M,\]
where $L_B$ and $L_C$ are the loss rates of $B$ and $C$ upon their defaults,
and $\l_M\geq 0$ is market-wide funding spread.
We may treat $\l_M$ as an indicator of
market-wide funding liquidity. To some extent $\l_M$, can be used to address the issue of liquidity valuation adjustment (LVA).

There is also a possibility that the risk-less value of the derivative has taken into account the
cost created by the repo spread, meaning that the risk-less value of the derivative is instead
\[
\hat V_s(0)\defn E^{Q_s}_0[\hat V_T],\]
where ${\Bbb Q}_s$ is equivalent to ${\Bbb Q}$ so that under ${\Bbb Q}_s$ the price of the underlying asset adjusted by the funding cost through repos,
\[
\td S_t=\hat S_te^{-\int^t_0\l_Sdu}={S_te^{\int^t_0(q_u-\l_S)du}\over B_t},\]
is a martingale. Obviously, this new measure, ${\Bbb Q}_s$, is defined by
\[
\left . {d{\Bbb Q}_s\over d{\Bbb Q}}\right|_{{\cal F}_t}=e^{\int^t_0-{1\over 2}\g^2 du+\g dW_u},\]
with
\[
\g=-{\l_S\over \s_S}. \]
It is straightforward to show the following relationship between the two version of risk-less values:
\begin{equation*}
\begin{split}
\hat V_s(0)=\hat V_e(\td S_0,0).
\end{split}
\end{equation*}
In terms of $V_s(0)=\hat V_s(0)$ and ${\Bbb Q}_s$, we have a simpler expression for the fair value $V_0$.
\begin{prop}\label{p23}
The value of a derivative under stochastic credit, debit and funding risks is given by
\begin{equation*}
\begin{split}
V_0=V_s(0)&+E^{Q_s}_0[1_{\{\tau=\tau_C\leq T\}}(\hat V(\tau_C+)-\td V_s(\tau_C))] \\
&+E^{Q_s}_0[1_{\{\tau=\tau_B\leq T\}}(\hat V(\tau_B+)-\td V_s(\tau_B))] \\
&+E^{Q_s}_0\left[\int^{T\wedge\tau}_0\left(-x_B[\hat \b_B(u)-\hat c_u]^-+x_C[\hat \b_C(u)+\hat c_u]^-\right)du\right]\quad\Box
\end{split}
\end{equation*}
\end{prop}


\section{Evaluation of the Adjustment Values}
For a collateralized trade, the payment upon a default is described in (\ref{e2011}), with the values of the collaterals
given by either (\ref{e2012}) or (\ref{e2013}), where the MTM value of the derivative can take either
\begin{enumerate}
\item the risk-free value, $M(\tau)=V_e(\tau)$, or
\item the pre-default value, $M(\tau)=V(\tau)$.
\end{enumerate}

For simplicity or analytical tractability in the valuation of CVA and FVA, we assume hereafter that 1) the hazard rates for defaults are
deterministic and 2) the value of the collateral does not exceed the MTM value of the derivatives, i.e. $|c(\tau)|\leq |M(\tau)|$.
When numerical methods are adopted for the valuation, these assumptions are not necessary.

\subsection{When $M(\tau)=V_e(\tau)$}
When the MTM value takes the risk-free value, the loss upon the default of $B$ is
\begin{equation}
\begin{split}\label{e3120}
V(\tau_B+)- V_e(\tau_B)&=c^+(\tau_B)+ V^-_e(\tau_B)-  V_e(\tau_B) \\
&=-V_e^+(\tau_B)+R_B\left(V_e(\tau_B)-H+X\right)1_{\{V_e(\tau_B)>H\}},
\end{split}
\end{equation}
where we have made use of the identity
\[
 f_t= f_t^++ f_t^-\]
for general real functions.
Similarly we also have the following general expression of default payment by $C$
\begin{equation}
\begin{split}\label{e3121}
  V(\tau_C+)-  V_e(\tau_C)&=-V^-_e(\tau_C)+R_C\left(V_e(\tau_C)+H-X\right)1_{\{V_e(\tau_C)\leq -H\}}.
\end{split}
\end{equation}
Corresponding to cases with or without collateral, we have 1) $R_B=R_C=1$ for $H>0$ or 2) $0\leq R_B,R_C <1$ for $H=X=0$.

\subsubsection{Evaluation of CVAs}
Plugging (\ref{e3120}) and (\ref{e3121}) to the CVA terms in (\ref{e2190}), we obtain
\begin{equation*}
\begin{split}
{CVA}_B
&=-E^{Q}_0\left[1_{\{\tau=\tau_B\leq T\}}\left({CC}( S_0,\tau_B,0,0)-R_B{CC}( S_0,\tau_B, H, X)\right)\right],\\
 {CVA}_C
&=-E^{Q}_0\left[1_{\{\tau=\tau_C\leq T\}} \left({CP}( S_0,\tau_C,0,0)-R_C{CP}( S_0,\tau_C, H, X)\right)\right],
\end{split}
\end{equation*}
where
\begin{equation*}
\begin{split}
& {CC}( S_0,\tau_B, H, X)=E^{Q}_0\left[\left(\hat V_e(\tau_B)-\hat H\right)^++\hat X1_{\{\hat V_e(\tau_B)>\hat H\}}\right], \\
& {CP}(S_0,\tau_C, H, X)=E^{Q}_0\left[\left(\hat V_e(\tau_C)+\hat H\right)^--\hat X1_{\{\hat V_e(\tau_C)\leq -\hat H\}}\right].
\end{split}
\end{equation*}
Note that both ${CC}$ and ${CP}$ can be treated as the discount values of a usual compound option plus a digital option.
For general payoff functions, the compound plus digital options can be conveniently evaluated by numerical methods, including finite difference methods and Monte Carlo simulations.
For vanilla call or put options, there are closed-form solutions for the compound options (Geske, 1979), which are provided in the appendix for completeness.

Since the first default of either party follows a Poisson process, we know that, conditional on ${\cal F}_0\vee\{\tau>0\}$, the probability density function of the first default by either party is
\begin{equation*}
\begin{split}
      f_i(u)=\l_i(u)e^{-\int^u_0(\l_B(v)+\l_C(v))dv},\quad \mbox{$i=B$ or $C$.}
\end{split}
\end{equation*}
So the evaluation of CVAs is just a matter of deterministic integrations:
\begin{equation*}
\begin{split}
&{CVA}_B 
=-\int^T_0f_B(u)\left({CC}( S_0,u,0,0)-R_B{CC}( S_0,u, H, X)\right) du, \\
&{CVA}_C 
=-\int^T_0f_C(u)\left({CP}( S_0,u,0,0)-R_C{CP}( S_0,u, H, X)\right) du.
\end{split}
\end{equation*}

When a derivative is either assets ($ V(T)\geq 0$) or liabilities ($ V(T)\leq 0$), there are
\begin{equation*}
\begin{split}
&{CC}( S_0,\tau_B,0,0)= V_e^+(0), \\
&{CP}( S_0,\tau_C,0,0)= V_e^-(0).
\end{split}
\end{equation*}
If, furthermore, if the values of the collaterals are proportional to the risk-less value, corresponding to $H=X=0$, then the evaluation
of the ${CVA}_B$ and ${CVA}_C$ can be much simplified:
\begin{equation*}
\begin{split}
 {CVA}_B
&=-E^{Q}_0\left[1_{\{\tau=\tau_B\leq T\}}\right](1-R_B) V^+_e(0)\\
 {CVA}_C
&=-E^{Q}_0\left[1_{\{\tau=\tau_C\leq T\}}\right](1-R_C) V^-_e(0).
\end{split}
\end{equation*}
Let $L_B=1-R_B$ and $L_C=1-R_C$, which can be treated as the loss rates. By working out the default probabilities we arrive at
\begin{prop} 
When the derivative is either an asset or a liability and both default intensities and loss rates are constants, there are
\begin{equation*}
\begin{split}
CVA_B 
&={-\l_B\over \l_B+\l_C}\left[1-e^{-(\l_B+\l_C)T}\right] L_B V^+_e(0), \\
CVA_C 
&={-\l_C\over \l_B+\l_C}\left[1-e^{-(\l_B+\l_C)T}\right] L_C V^-_e(0). \quad \Box
\end{split}
\end{equation*}
\end{prop}

The above results simply show that a $CVA$ is equal to the product of first-default probability of the party of liability and the lost value to the surviving party, which generalize the existing result of
CVA under unilateral default risk of the counterparty (see e.g. Pyktin and Zhu (2007) or Gregory (2009)):
\begin{equation*}
\begin{split}
CVA_C 
&=-\left[1-e^{-\l_CT}\right] L_C V^-_e(0).
\end{split}
\end{equation*}

\subsubsection{Evaluation of FVA}
In general, the evaluation of FVA has to resort to numerical methods like lattice-tree methods, due to the presence of the derivative premium in the funding cost terms and the path-dependence nature of the margin costs. Denote

\begin{equation*}
\begin{split}
{FVA}_S=&E_0\left[\int^{T}_01_{u\leq\tau}\l_S\hat S_u\p_{\hat S}\hat V_u du\right], \\
{FVA}_B=&-E_0\left[\int^{T}_01_{u\leq\tau}x_B[\hat \b_B(u)-\hat c_u]^- du\right], \\
{FVA}_C=&E_0\left[\int^{T}_01_{u\leq\tau}x_C[\hat \b_C(u)+\hat c_u]^- du\right],
\end{split}
\end{equation*}
which can be evaluated by different strategies.

To evaluate $FVA_S$, we take advantage of the independence between default risks and the market risk, which leads to
\begin{equation*}
\begin{split}
{FVA}_S=&
\int^T_0\l_SE_0\left[\hat S_u\p_{\hat S}\hat V_u\right] E_0[1_{\tau\geq u}]du,
\end{split}
\end{equation*}
with
\[
E_0[1_{\tau\geq u}]=
e^{-\int^u_0(\l_B+\l_C)du}.\]
The only problem left for valuating ${FVA}_S$ is to calculate $E_0[\hat S_u\p_{\hat S}\hat V_u]$, the expected value of the delta hedging term, which however is not readily available. For many types of derivatives, there is however a shortcut. In fact, we can
approximate the true delta by that of the risk-free value, i.e., to let
$\p_{\hat S}\hat V_u\approx \p_{\hat S}\hat V_e(u)$, then, due to the martingale property of the delta hedging term for the risk-free derivative, we obtain
\begin{equation*}
\begin{split}
{FVA}_S 
\approx &\int^T_0\l_S E_0[1_{\tau\geq u}]du\ \hat S_0\p_{\hat S}\hat V_e(0).
\end{split}
\end{equation*}

$FVA_B$ and $FVA_C$ depend on $V_0$, since functions $\hat\b_B(t)$ and $\hat\b_C(t)$ in the integrand start from $\b_B(0)=V^+_0$ and $\b_C(0)=-V^-_0$. Once a value of $V_0$ is given, ${FVA}_B$ and ${FVA}_C$ can be conveniently evaluated by either Monte Carlo simulation methods or lattice tree methods, so they can be treated as functions of $V_0$. Therefore, $V_0$ satisfies the following
equation:
\begin{equation}
V_0-{FVA}_B(V_0)-{FVA}_C(V_0)=V_e(0)+{CVA}_B+{CVA}_C+FVA_S,
\end{equation}
where the right-hand side is a constant while the left-hand side is actually a monotonically increasing function of $V_0$, which
thus can be easily solved by a root-finding method like bisection method or fixed-point iteration method.

{\bf Example 1}. For demonstration, we price an at-the-money (ATM) European call option for a range of issuer's hazard rate. Let the stock price be $S_t=100$, with volatility $\s=20\%$ and dividend yield $q_t=0$,
and let the interest rate be $r_t=3\%$. When there is no credit and funding risks, the price of one-year ATM call option is $V_e=9.4134$,
according to the Black-Scholes formula. In the presence of counterparty default and funding risks, we need to make valuation adjustments to the risk-free value. Since the call option is a liability to the issuer, there are $CVA_C=FVA_C=0$, so that
\[
V_0=V_e(0)+CVA_B+FVA_S+FVA_B.\]
We calculate the option value and its valuation adjustments using a binomial tree model for the underlying share price, with the time-step size of $\D t=1/52$.

We take the following parameters for funding or credit risks. The repo basis is $\l_S=0.75\%$, the market-wide funding spread is $\l_M=0$, the loss rates are $L_B=L_C=0.6$, the default intensity for $C$ is $\l_C=1.5\%$, and we let $\l_B$, the default intensity for $B$, vary from 0 to 300 basis points.  We consider both no-CSA and CSA trades. For the no-CSA trade, we apply a 40\% recovery rate upon default to the risk-free value. For the CSA trade with cash collateral, we take the threshold value and the minimum transfer amount to be $H=4$ and $X=2$, respectively.

Figure 1 and 2 show the option values together with its valuation adjustments, without and with collaterals. In both figures, the plot on the right gives an enlarged view of the adjustment values only. As we can see, the corresponding funding valuation adjustment terms in both cases are very close in value, and they are insensitive to the hazard rate of $B$. In particular, the funding valuation adjustment due to margining is negligibly small (and is under $10^{-5}$). This can be explained as follows. Under our diffusion model for the underlying share, the call option can be perfectly replicated, until the first default or maturity, so the value of the margin account, $\b_B(t)$, equals to the value of the option. When the value of the cash collateral is smaller than the value of the option, the margin balance after collateral withdrawal will stay positive and thus incur no cost. The cost for delta hedging using repos is neither negligible nor sensitive to the hazard rate of $B$. The credit valuation adjustment, meanwhile, is sensitive to the hazard rate of $B$, with its magnitude depending on the amount of collateral being put down. In both cases, the CVAs are  insignificant when compared with the option values.

\centerline{\epsfxsize=6in \epsfbox{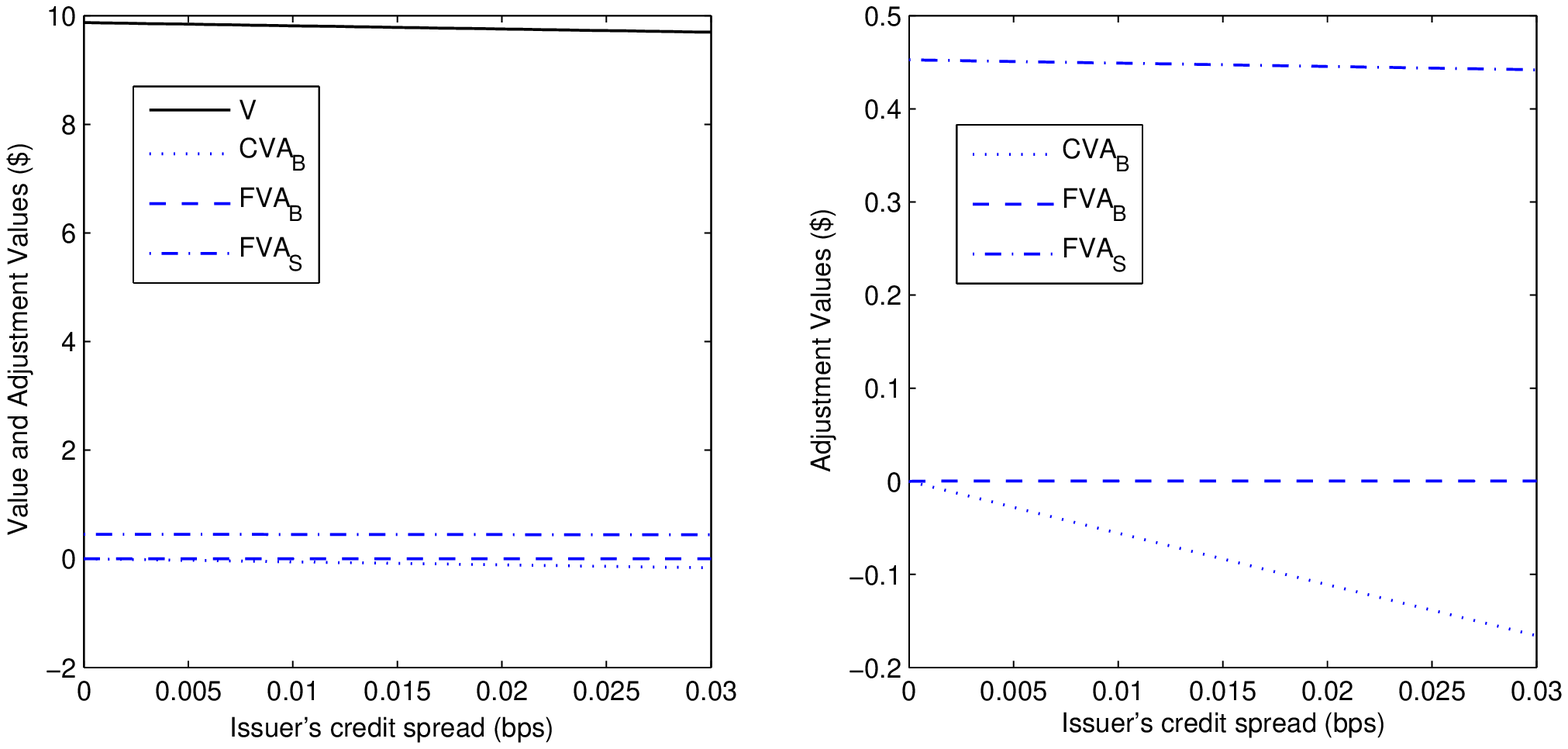}}
\centerline{{\bf Figure 1.} Derivatives value and the adjustments without collateral}

\newpage
\vspace*{0.0in}
\centerline{\epsfxsize=6in \epsfbox{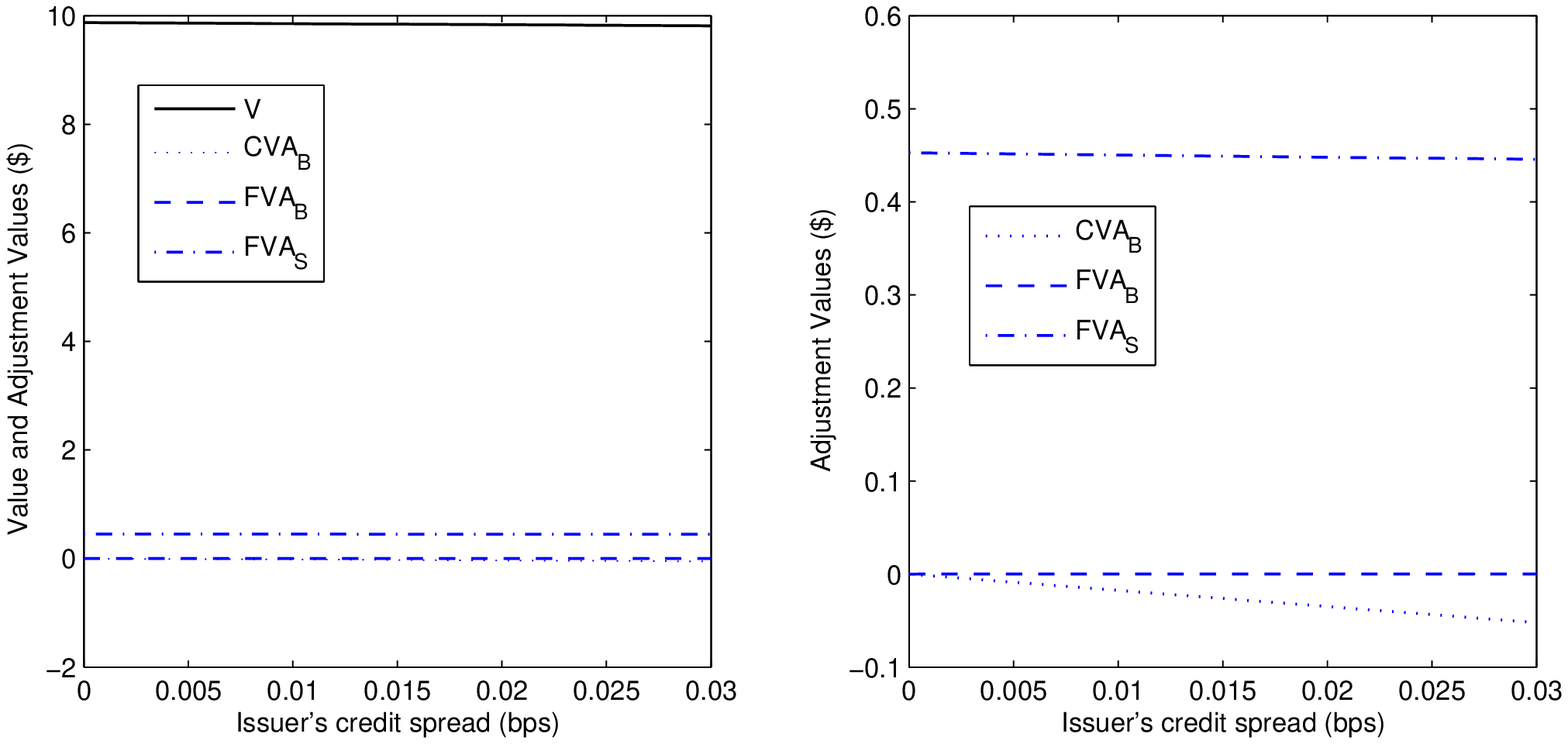}}
\centerline{{\bf Figure 2.} Derivatives value and the adjustments with cash collateral}


\subsection{When $M(\tau)=V(\tau)$}
When the MTM value upon a default takes the pre-default value, formula (\ref{e2190}) is not too useful, and we thus revisit (\ref{e2110}), the governing equation for the pre-default value. Given the value of the collateral being capped by the pre-default value, we have the jump
size be either
\begin{equation*}
\begin{split}
  \D \hat V_B=&
  \hat c^+_t-\hat V^+_t, \quad\mbox{or}\quad \D \hat V_C=\hat c^-_t-\hat V^-_t,
\end{split}
\end{equation*}
with $c_t$ given by (\ref{e2012}) or (\ref{e2013}).
Plug in the above expressions into (\ref{e2110}), we obtain
\begin{equation*}
\begin{split}
\p_t\hat V_t+{1\over 2}\s_S^2\hat S^2\p^2_{\hat S}\hat V_t
=&-\l_B\D \hat V_B(t)-\l_C\D \hat V_C(t)\\
&-\l_S\hat S\p_{\hat S}\hat V_t+x_B[\hat\b_B(t)-\hat c_t]^--x_C[\hat\b_C(t)+\hat c_t]^-,
\end{split}
\end{equation*}
augmented with either default payment (\ref{e2011}) or terminal payment at maturity.
This unconventional terminal-value problem can be very expensive to solve, while analytical methods seem less likely.

For illustration, let us consider the pricing of a derivative asset, with $\hat V(T)\geq 0$, in the absence of collateral and with proportional recovery upon default. It can be argued that there is $\hat \b_B(t)=\hat V_t\geq 0$ due to perfect replication. So, the
cost for margining reduces to zero and the unconventional terminal-value problem of the PDE can be simplified into
\begin{equation}
\begin{split}\label{e3290}
&\p_t\hat V_t+{1\over 2}\s_S^2\hat S^2\p^2_S\hat V_t=\l_BE_t[L_B]\hat V_t-\l_S\hat S\p_t\hat V_t,\quad 0\leq t\leq \tau_B\wedge T,\\
&\hat V(\tau_B+)=R_B\hat V(\tau_B),\quad\mbox{or}\quad \hat V(T)=\hat f(S),
\end{split}
\end{equation}
where $f(S)$ is the payoff at maturity. The solution to (\ref{e3290}) can be expressed using ${\Bbb Q}_s$ expectations:
\begin{equation}
\begin{split}\label{e3300}
\hat V_0=E^{Q_s}_0&\left[e^{-\int^{\tau_B}_0\l_BE_u[L_B]du}R_B\hat V(\tau_B)1_{\tau_B\leq T}\right] \\
&\qquad\qquad +E^{Q_s}_0\left[e^{-\int^T_0\l_BE_u[L_B]du}\hat f(S_T)1_{\tau_B>T}\right],
\end{split}
\end{equation}
which is an integral equation for $\hat V_0$, and it may only be solved recursively through a  time-stepping discretization method for PDEs.
Note that conditional on $\tau_B>T$, the first term of (\ref{e3300}) vanishes and solution
actually reduces to Equation (3) of Piterbarg (2010), when there is only funding cost.

\section{CVA and FVA for Interest-rate Swaps}

Next, we consider valuation adjustments to vanilla interest-rate swaps, the most liquid interest-rate derivatives traded over the counter. There will be two major differences from the theory for equity derivatives pricing established in section 2. First, there is no delta hedging for the swaps. Second, the cash collateral is revised in discrete time, which is the reality. In fact, with or without collateral, we can always evaluate the CVA terms analytically, using the swap market model. When there is cash collateral, the costs of margining are path-dependent, and their valuations should resort to
Monte Carlo simulations. To avoid getting into discussion of choice of term structure models, we in this article limit ourselves
to the evaluation of the credit valuation adjustment in the absence of collateral, when there is no margining cost.
For simplicity, we take the same payment frequency for both fixed and floating legs.

Our swap pricing will be based on the swap market model for swap and swaption pricing, which is briefly introduced below. Let
$P(t,T)$ be the OIS discount curve, $f_j(T_j)$ be the LIBOR rate for the period $(T_j,T_{j+1})$, ${\Bbb Q}_{j+1}$ be the $T_{j+1}$-forward measure corresponding to $P(t,T_{j+1})$ as numeraire. Define
\begin{equation*}
\begin{split}
A_{m,n}(t)&=\sum^{n-1}_{j=m}\D T_jP(t,T_{j+1}), \\
s_{m,n}(t)&=\sum^{n-1}_{j=m}{\D T_jP(t,T_{j+1})\over A_{m,n}(t)}E^{Q_{j+1}}_t[f_j(T_j)],
\end{split}
\end{equation*}
for $t\leq T_m$. Note that after the 2007-08 financial tsunami, LIBOR rates are no longer considered riskless and they are no longer martingales under their cash-flow measures. Without credit and funding risks, the values of a vanilla fixed-for-floating payer's swap of tenor $(T_m,T_n)$ is given by
\begin{equation*}
\begin{split}
V_e(t)=&A_{m,n}(t)(s_{m,n}(t)-s).
\end{split}
\end{equation*}
Under the swap market model, the values of call and put options on fixed-for-floating swap are given by
\begin{equation}
\begin{split}\label{e4200}
BC(t,T_m,s_{m,n}(t),s,m,n,\s)&=A_{m,n}(t)(s_{m,n}(t)\Phi(d_1)-s\Phi(d_2)), \\
BP(t,T_m,s_{m,n}(t),s,m,n,\s)&=A_{m,n}(t)(s\Phi(-d_2)-s_{m,n}(t)\Phi(-d_1)),
\end{split}
\end{equation}
where $\Phi(\cdot)$ is the standard normal accumulative function, and
\[
d_{1,2}={\ln{s_{m,n}(t)\over s}\pm{1\over 2}\s^2(T_m-t)\over \s\sqrt{T_m-t}}.\]
For derivation of the swap and swaption formulae we refer to, e.g., Wu (2009).

We now proceed to evaluate $DVA$.
According to the rules of default settlement, there is
\begin{equation}
\begin{split}\label{e3470}
{DVA}=&E_0\left[1_{\{\tau=\tau_B<T\}}(R_B\hat V^+_e(\tau_B)-\hat V^+_e(\tau_B))\right] \\
=&(R_B-1)\sum^n_{j=1}E_0\left[\hat V^+_e(T_j)\right]E_0\left[1_{T_{j-1}\leq\tau=\tau_B\leq T_j}\right],
\end{split}
\end{equation}
where
\begin{equation}\label{e3480}
E_0[1_{T_{j-1}\leq \tau=\tau_B\leq T_j}]={e^{-\int^{T_{j-1}}_0(\l_B+\l_C)du}}(1-e^{-\int^{T_j}_{T_{j-1}}\l_Bdu}).
\end{equation}
Since
\begin{equation*}
\begin{split}
\hat V^+_e(T_j)=&\hat A_{j,n}(T_j)(s_{j,n}(T_j)-s)^+
\end{split}
\end{equation*}
is the payoff of a swaption, we have
\begin{equation}
\begin{split}\label{e3490}
&E_0\left[\hat V^+_e(T_j)\right]={BC}(0,T_j,s_{j,n}(0),s,j,n,\s). 
\end{split}
\end{equation}
Putting (\ref{e3480}) and (\ref{e3490}) back to (\ref{e3470}), we obtain $DVA$.

The formulae for $CVA$ can be derived similarly, and it is
\begin{equation*}
\begin{split}
{CVA}
=&(R_C-1)\sum^{n}_{j=1}E_0\left[\hat V^-_e(T_j)\right]E_0[1_{T_{j-1}\leq\tau=\tau_C\leq T_j}],
\end{split}
\end{equation*}
with
\begin{equation*}
\begin{split}
E_0[1_{T_{j-1}\leq\tau=\tau_C\leq T_j}]=
{e^{-\int^{T_{j-1}}_0(\l_B+\l_C)du}}(1-e^{-\int^{T_j}_{T_{j-1}}\l_Cdu})
\end{split}
\end{equation*}
and
\begin{equation*}
\begin{split}
E_0\left[\hat V^-_e(T_j)\right]=&-{BP}(0,T_{j},s_{j,n}(0),s,j,n,\s).
\end{split}
\end{equation*}

According to Proposition 1, the value of the swaps exposed to counterparty default risk and funding risk is
\begin{equation*}
\begin{split}
\hat V_0=&\hat V_e(0)+{DVA}+{CVA}.
\end{split}
\end{equation*}
The fair swap rate after adjusted for counterparty default risks can be solved from $\hat V_0=0$.

{\bf Example 2}. We consider the pricing of a ten-year vanilla swap with semi-annual payment frequency, such that $\D T_j=0.5$. The OIS discount curve and Euribor forward-rate curve are constructed using data of August 24, 2012, and they are shown in Figure 3.
We take $s=s_{0,20}(0)=1.45\%$, the risk-free ATM swap rate so that $V_e(0)=0$.
For the adjustment terms, we take
\begin{equation*}
\begin{split}
&R_B=R_C=0.4, \\
&\l_M=0, \quad \l_C=0.015\%,\quad\mbox{and}\quad \l_B=0.000:0.005:0.03.
\end{split}
\end{equation*}
Figure 4 displays the swap value together with the adjustment terms. Without surprise, it shows that $DVA$ decreases with an increasing $\l_B$, while
$CVA$ remains close to a constant due to the constant hazard rate for $C$.

\vspace*{0.25in}
\centerline{\epsfxsize=6in \epsfbox{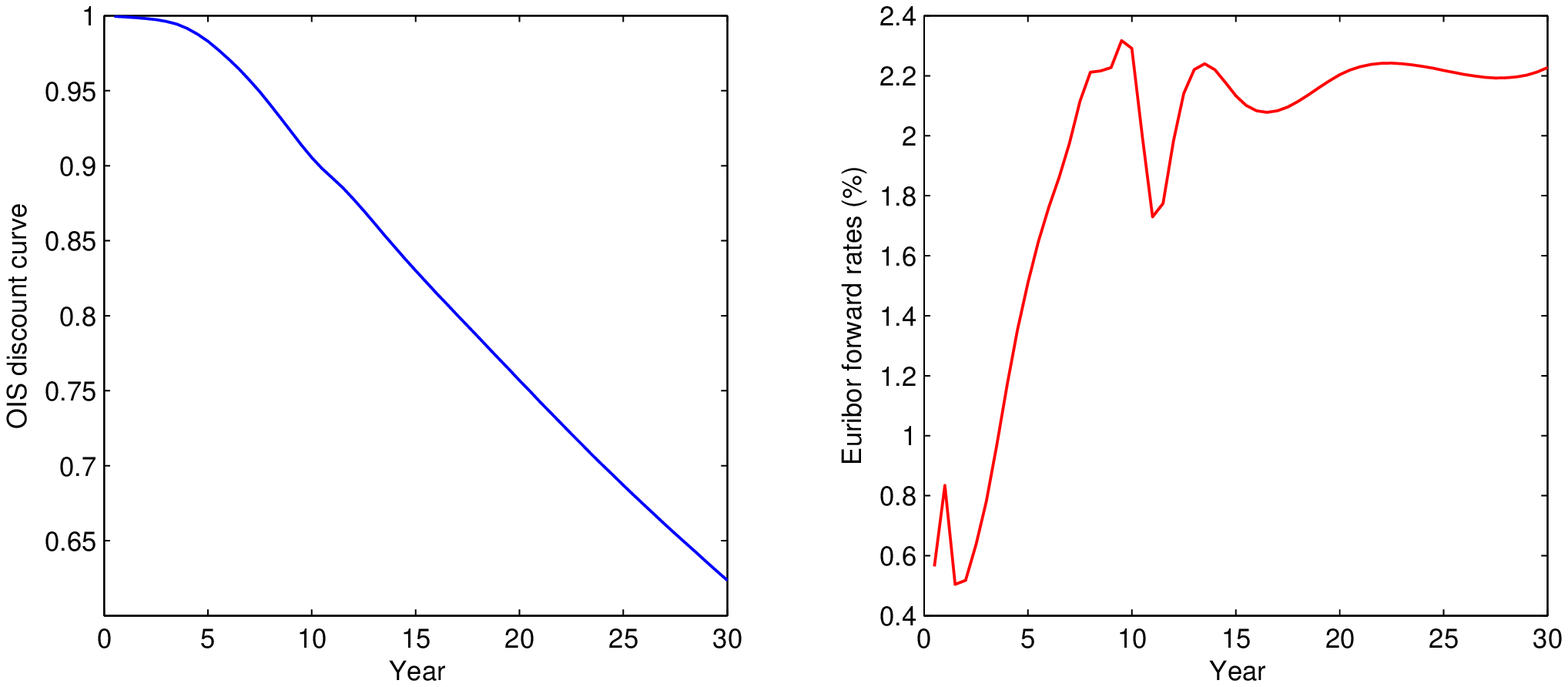}}
\centerline{{\bf Figure 3} OIS discount curve and Euribor forward rates}

\vspace*{0.25in}
\centerline{\epsfxsize=4in \epsfbox{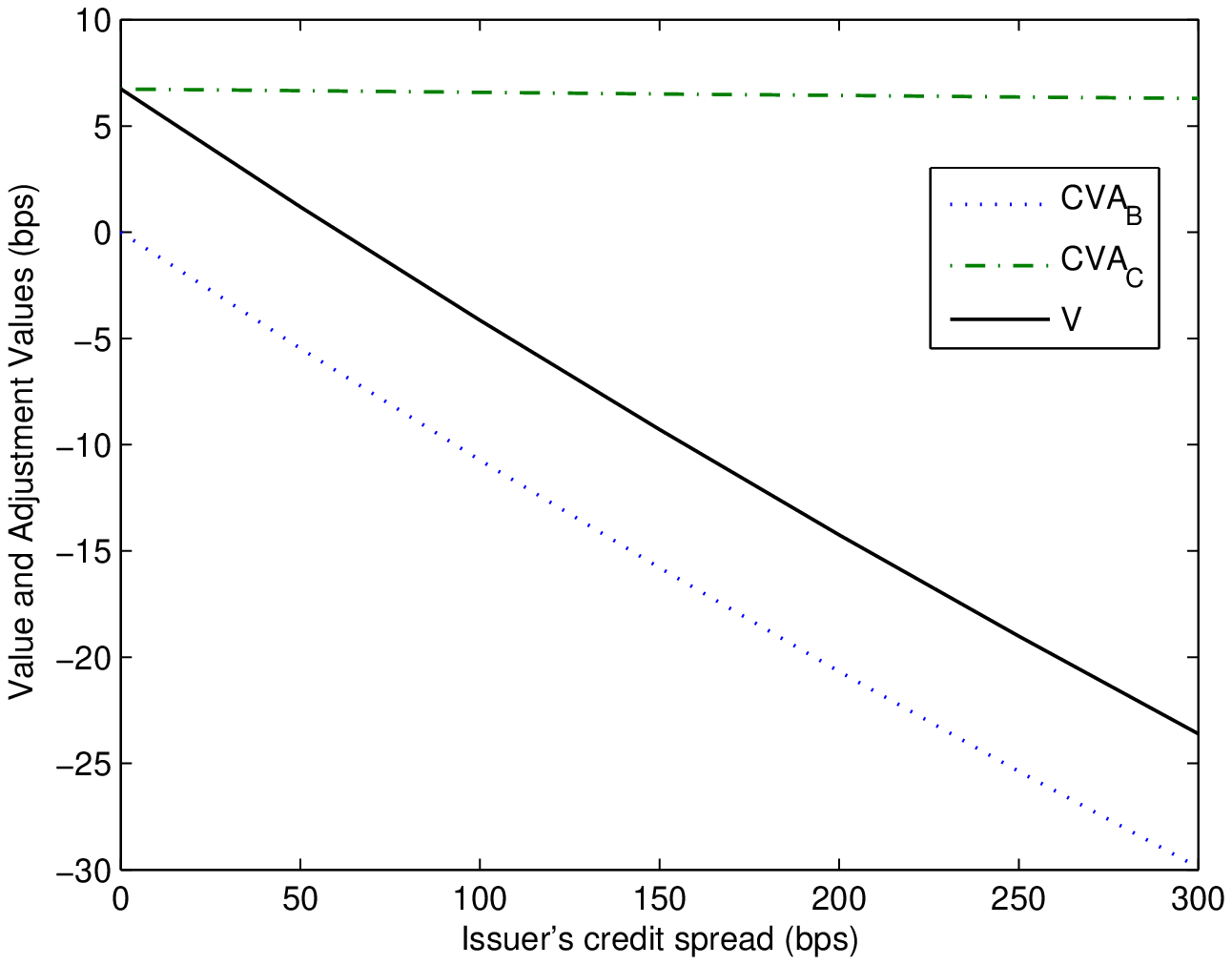}}
\centerline{{\bf Figure 4} Credit valuation adjustments for $B$ and $C$}

\section{Conclusions}
We combine replication pricing with expectation pricing to derive the CVA and FVA of a derivative security subject to bilateral default and funding risks. While the evaluation of CVA can often be done analytically or semi-analytically, the evaluation of FVA in general should resort to numerical methods like Monte Carlo simulations. The results can be directly generalized to portfolios of derivatives with collaterals and/or netting agreements, and the framework can be applied to deal with re-hypothecation. The combined approach can be easily generalized to accommodate more general asset-price dynamics like dynamics like
jump-diffusion processes.

There is room to enhance the results of this article. The analysis of this article is based on the assumption of independence between the credit risks of the two parties and the dynamics of the underlying asset price. This assumption must be removed if we want to deal with the so-called wrong-way risk. Also, re-hypothecation is a
current issue and may possibly be addressed under the framework of this article. Finally, the adjustment terms derived in the article are for European options. For application purposes, we need to generalize the results of this article to American or Bermudan options.

\appendix
\section{Price formulae for compound plus digital options}

The compound-call plus digital option we encounter in the article can be expressed as
\begin{equation}
\begin{split}\label{a010}
{CC}(S_0,u, H, X)=&E^{Q}_0\left[\left(\hat V_e(u)-\hat H+\hat X\right)1_{\{\hat V_e(u)>\hat H\}}\right] \\
=& E^{Q}_0\left[\left(\hat V_e(u)-\hat H\right)^+\right]+E^{Q}_0\left[\hat X1_{\{\hat V_e(u)>\hat H\}}\right],
\end{split}
\end{equation}
where the first term is a standard compound call option. Under stochastic interest rate, the compound call option can be treated as follows
\begin{equation}
\begin{split}\label{a020}
E^{Q}_0\left[\left(\hat V_e(u)-\hat H\right)^+\right]
=&{P(0,u)}E^{Q}_0\left[{P(u,u)\over P(0,u) B_{u}}\left(V_e(u)-H\right)^+\right] \\
=&{P(0,u)}E^{Q_{u}}_0\left[\left(V_e(u)-H\right)^+\right].
\end{split}
\end{equation}
Here, ${\Bbb Q}_{u}$ is the $u$-forward measure corresponding to numeraire $P(t,u)$, the $u$-maturity risk-free discount bond. Compound options like (\ref{a020}) can be evaluated numerically or, in some circumstance, analytically.

When the derivatives underlying the compound option is a usual call or put option, we know from Geske (1979) that the value of compound options can be obtained in closed forms. Let $C(S_{u},u;K,T)$ denote the time-$u$ price of a call option with maturity $T$ and strike $K$, then, assuming constant dividend yield, the European call-on-call option is
\begin{equation}
\begin{split}\label{a030}
E^{Q_{u}}_0\left[\left(C(S_{u},u;K,T)-H\right)^+\right]
=&{S_te^{-qT}\over P(0,u)}\Phi_2(a_+,b_+,\sqrt{u/ T}) \\
-{KP(0,T)\over P(0,u)}&\Phi_2(a_-,b_-,\sqrt{u/ T})
-{H}\Phi(a_-),
\end{split}
\end{equation}
where $\Phi_2(x,y;\rho)$ is the two-dimensional cumulative normal distribution functions,
\begin{equation*}
\begin{split}
&a_+={\ln{S_0e^{-qu}\over S_u^*}+{1\over 2}\s^2u\over \s\sqrt{u}},\qquad a_-=a_+-\s\sqrt{u}, \\
&b_+={\ln{S_0e^{-qT}\over K}+{1\over 2}\s^2T\over \s\sqrt{T}},\qquad b_-=b_+-\s\sqrt{T},
\end{split}
\end{equation*}
$S_{u}^*$ solves $C(S_{u},u;K,T)=H$, and $\s$ is the volatility of ${S_te^{-qT}\over P(0,u)}$, the $u$-forward price of the underlying. There is also
\begin{equation}
\begin{split}\label{a040}
E^{Q}_0\left[\hat X1_{\{\hat C(u)>\hat H\}}\right]={P(0,u)}E^{Q_u}_0\left[X1_{\{C(u)>\hat H\}}\right]={P(0,u)}X\Phi(a_-).
\end{split}
\end{equation}
Combining (\ref{a010}) to (\ref{a040}), we obtain the price formulae for the compound-call-plus-digital option:
\begin{equation*}
\begin{split}
{CC}(S_0,u, H, X)
=&{S_0e^{-qT}}\Phi_2(a_+,b_+,\sqrt{u/ T}) \\
-{KP(0,T)}&\Phi_2(a_-,b_-,\sqrt{u/ T})
-{(H-X)P(0,u)}\Phi(a_-),
\end{split}
\end{equation*}

Function ${CP}(S_0,\tau_C, H, X)$ may be nonzero only if $\hat V_e(\tau_C)\leq 0$, when the derivative is a liability to $C$. We have
\begin{equation*}
\begin{split}
& {CP}(S_0,\tau_C, H, X)=-E^{Q}_0\left[\left(-\hat V_e(\tau_C)-\hat H+\hat X\right)1_{\{-\hat V_e(\tau_C)\geq \hat H\}}\right],
\end{split}
\end{equation*}
which can be evaluated in the same way as for ${CC}(S_0,u, H, X)\quad\Box$

\end{document}